\documentclass[manuscript=article]{achemso}
\setkeys{acs}{maxauthors=60}
\setkeys{acs}{etalmode=truncate}
\setkeys{acs}{doi=true}
\usepackage[T1]{fontenc}

\usepackage{geometry}
\usepackage{setspace}

\usepackage{achemso}

\usepackage{graphicx}
\usepackage{float}
\newfloat{scheme}{htbp}{los}
\floatname{scheme}{Scheme}
\floatname{chart}{Chart}
\newfloat{graph}{htbp}{loh}
\DeclareUnicodeCharacter{2212}{-}

\usepackage{chemformula} % Formulas using \ch{}
\usepackage[version = 4]{mhchem} % Formulas using \ce{}
\usepackage{times}
\usepackage{dcolumn}% Align table columns on the decimal point
\usepackage{bm}% bold math
\usepackage{amsmath}
\usepackage[colorlinks=true, allcolors=blue]{hyperref}
\usepackage{ulem} % strike out 
\usepackage{chngcntr} % numbering for appendix
\usepackage{enumitem}
\usepackage{lineno}
\usepackage{adjustbox}
\usepackage{xr}
\usepackage{cleveref}
\author{Hyo-sun Jin}
\affiliation{Department of Physics Education, Kyungpook National University, Daegu 41566, South Korea}
\alsoaffiliation{KNU G-LAMP Project Group, KNU Institute of Basic Sciences, Kyungpook National University, Daegu 41566, South Korea}
\author{Jiyeon Kim}
\affiliation{Department of Physics Education, Kyungpook National University, Daegu 41566, South Korea}
\author{Sooran Kim}
\email{sooran@knu.ac.kr}
\affiliation{Department of Physics Education, Kyungpook National University, Daegu 41566, South Korea}
\alsoaffiliation{KNU G-LAMP Project Group, KNU Institute of Basic Sciences, Kyungpook National University, Daegu 41566, South Korea}
\title{Electronic Structure Descriptors for CO$_2$ Conversion Activity in Perovskite Oxides}

\begin{document}

\maketitle

\begin{abstract}
  Electronic structure descriptors have been widely used to rationalize catalytic activity, but their application to CO$_2$-to-CH$_4$ conversion in perovskite oxides remains relatively limited. Here, we investigate six electronic structure descriptors related to band centers and bandwidths. We find that the unoccupied transition-metal (TM) $3d$-band center, charge-transfer energy, and the ratio of the unoccupied TM $3d$ to occupied O $2p$ bandwidths, $W_{3d}/W_{2p}$, exhibit negative correlations with CH$_4$ activity. These qualitative trends are further quantified using a parametric brute-force searching (BFS) approach to obtain explicit fitting equations. Notably, the selected equations depend only on $W_{3d}/W_{2p}$, providing an electronic structure perspective on the enhanced activity in double perovskites.
  We apply the selected equation to screen 24 La-based double perovskites. This screening suggests that La$_2$CoGaO$_6$, La$_2$CoAlO$_6$, and Ti-containing compositions are promising candidates. The present approach can help identify descriptor-activity relationships and guide materials screening in perovskite oxides.
\end{abstract}
\newpage

\begin{figure}[b]
\centering
\includegraphics[width=0.55\columnwidth]{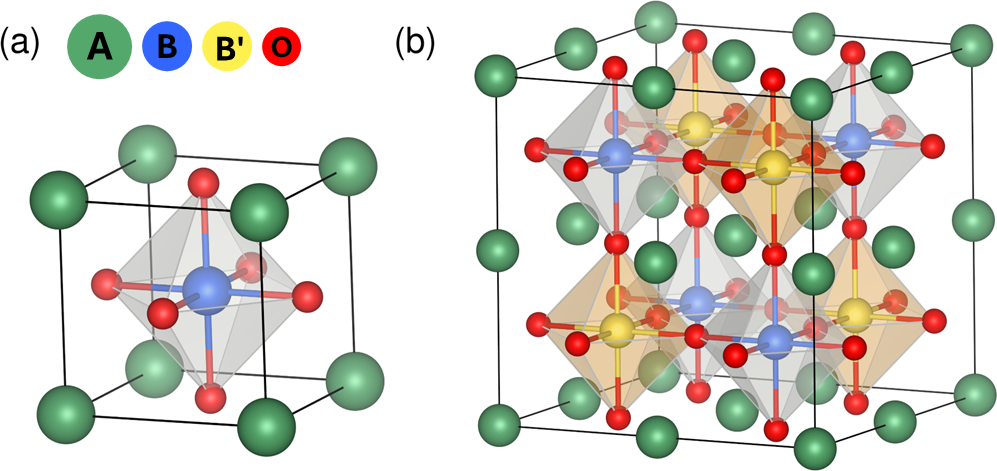}
\caption{\label{fig:Fig1}Crystal structures of (a) a simple perovskite (ABO$_3$) and (b) a double perovskite (A$_2$BB$^\prime$O$_6$). In the double perovskite structure, two distinct metal ions (B and B$^\prime$) are located at the octahedral sites.}
\end{figure}

 The electrochemical CO$_2$ reduction reaction (CO$_2$RR) has attracted considerable interest as a promising route for converting an abundant greenhouse gas into value-added chemicals and fuels \cite{qiao_review_2014, wang_electrocatalysis_2021, overa_electrochemical_2022, li_rational_2022, nitopi_progress_2019}. However, achieving efficient CO$_2$RR remains challenging because of the high chemical stability of the linear CO$_2$ molecule, the presence of multiple CO$_2$ reduction pathways, and the competing hydrogen evolution reaction (HER) \cite{popovic_stability_2020, birdja_advances_2019, saha_selectivity_2022, liu_understanding_2017, wang_rational_2023}. Consequently, developing catalysts that exhibit high activity, selectivity, and stability is essential for CO$_2$RR catalysis \cite{tan_recent_2021, shi_barriers_2016, zhong_accelerated_2020, wei_electronic_2023}.

 Noble-metal catalysts have shown excellent electrocatalytic performance for CO$_2$RR \cite{sun_rational_2020, kauffman_experimental_2012, zhu_shapecontrolled_2019}, but their scarcity and high cost limit their practical applications. Perovskite oxides have therefore emerged as attractive alternatives because their compositions and electronic structures can be systematically tuned within a structurally robust metal–oxygen framework \cite{royer_perovskites_2014}. Simple perovskites adopt the ABO$_3$ structure, while double perovskites can be described as A$_2$BB$^\prime$O$_6$, where two distinct metal ions occupy the octahedral sites, as shown in Fig. \ref{fig:Fig1}. Their broad compositional flexibility enables systematic control over transition-metal oxidation states, TM–O hybridization, charge-transfer characteristics, and oxygen nonstoichiometry. Oxygen vacancies can modify surface adsorption, charge redistribution, and electron-transfer processes \cite{zhao_structural_2024, zhang_perovskite_2021}. Double perovskites further provide an additional degree of compositional freedom through the combination of two different B-site cations, and previous studies have reported enhanced CO formation in multi-cation perovskite systems relative to their parent or simple perovskite counterparts \cite{pomiro_co2_2023, lim_fundamental_2021, goto_enhanced_2024}.

 To rationalize catalytic activity trends from electronic structure, various descriptors have been proposed, including the \textit{d}-band center, \textit{d}-band shape, $e_g$ occupancy, O $2p$-band center, and charge-transfer energy \cite{hammer_electronic_1995, hammer_theoretical_2000, suntivich_perovskite_2011, pbandcenter_grimaud, hong_CTE, xin_effects_2014}. Hammer and Nørskov established the \textit{d}-band model for understanding adsorption energetics and surface reactivity on transition-metal surfaces, highlighting the role of the \textit{d}-band center, $\varepsilon_d$, in controlling adsorbate-metal interactions \cite{hammer_electronic_1995, hammer_theoretical_2000}. This concept was further refined by considering $d$-band width and shape; for example, Vojvodic \textit{et al}. proposed a bandwidth-corrected descriptor, $\varepsilon_d+W_d/2$ \cite{vojvodic_electronic_2014}, while Xin \textit{et al}. suggested that the $d$-band shape, particularly the upper $d$-band edge, can describe adsorption trends on transition-metal alloys \cite{xin_effects_2014}.

For the oxygen evolution reaction (OER), electronic structure descriptors have also been actively developed for perovskite oxides \cite{suntivich_perovskite_2011, pbandcenter_grimaud, hong_CTE, yamada_2018, hong_descriptors_2016, jacobs_assessing_2019}. Suntivich \textit{et al}. correlated the OER activity of perovskite oxides with the surface transition-metal $e_g$ occupancy, identifying an optimal $e_g$ filling close to unity \cite{suntivich_perovskite_2011}. Later, the O $2p$-band center relative to the Fermi level was proposed as another important descriptor for perovskite OER activity \cite{pbandcenter_grimaud, jacobs_assessing_2019}. Grimaud \textit{et al}. showed that an upward shift of the O $2p$-band center enhances metal-oxygen hybridization and OER activity, whereas an excessively high O $2p$-band center can reduce structural stability under OER conditions\cite{pbandcenter_grimaud}. Hong \textit{et al}. further reported that the charge-transfer energy, which reflects the energy separation between TM $3d$ and O $2p$ states, is closely associated with interfacial properties and OER mechanistic pathways \cite{hong_CTE}. Smaller charge-transfer energy generally correlates with higher OER activity in perovskite oxides \cite{hong_CTE,yamada_2018}.

 Descriptor-based approaches have also been applied to CO$_2$ activation and CO$_2$RR\cite{ma_machine-learning-augmented_2015,abraham_machine_2023,hwang_2019}. Ma \textit{et al}. developed a machine-learning model for CO$_2$ electroreduction catalyst screening on multimetallic alloys and suggested that descriptors related to the $d$-band center and $d$-band shape, together with local electronegativity, are important for describing CO adsorption energetics on alloy surfaces \cite{ma_machine-learning-augmented_2015}. For MXene catalysts, Abraham \textit{et al}. employed machine learning to investigate CO$_2$ activation on two-dimensional transition-metal carbides and nitrides and showed that the $d$-band center, surface metal electronegativity, and metal valence electron number are important descriptors for predicting CO$_2$ activation behavior \cite{abraham_machine_2023}. For perovskite oxides, Hwang \textit{et al}. combined experiments and DFT calculations to investigate CO$_2$RR and proposed the bulk O $2p$-band center as a useful descriptor for describing CH$_4$ activity trends. They showed that CH$_4$ activity follows a volcano-type dependence on the O $2p$-band center, with LaCoO$_3$ located near the optimum region \cite{hwang_2019}. Although various electronic descriptors have been extensively examined for OER in perovskite oxides, descriptor-based understanding of CO$_2$ conversion activity in perovskite oxides remains relatively limited. 

In this work, motivated by previous descriptor-based studies of surface reactivity and oxide electrocatalysis, we investigate band center, bandwidth-related, and charge-transfer energy descriptors for CO$_2$-to-CH$_4$ activity in perovskite oxides. We compare these descriptors with the experimentally reported CH$_4$ activities and examine the resulting correlations. To quantify the observed qualitative trends, we employ a parametric brute-force searching (BFS) approach to systematically explore explicit equations relating these descriptors to CH$_4$ activity. Through this analysis, we identify the ratio of the unoccupied TM $3d$ bandwidth to the occupied O $2p$ bandwidth as a descriptor for capturing the activity trends of simple and double perovskites. Finally, the selected expression is applied to screen 24 La-based double perovskites, highlighting La$_2$CoAlO$_6$, La$_2$CoGaO$_6$, and Ti-containing compositions with relatively high predicted activities.

Spin-polarized density functional theory (DFT) calculations were performed for a total of 32 perovskite oxides, comprising 8 simple and 24 double perovskite oxides. All calculations were carried out using the Perdew–Burke–Ernzerhof generalized gradient approximation (PBE-GGA) for the exchange-correlation functional \cite{gga}, as implemented in the projector augmented-wave method (PAW) within the Vienna ab initio Simulation Package (VASP) \cite{vasp_1994, vasp_1996}. 
To properly describe the strong correlation effects of transition-metal $d$ orbitals, the GGA+$U$ approach was employed with effective $U$ values of 3.0 eV for Ti, 3.5 eV for V and Cr, 4.0 eV for Mn, Fe, and Cu, 3.3 eV for Co, and 6.4 eV for Ni, as reported in previous studies \cite{Ti_u, suntivich_2014, vanadium_2006, vanadium_u2, Cu_u}. For rare-earth $f$ electrons, $U=8$ eV and $J=1$ eV were applied to Nd, while $U=9$ eV and $J=1$ eV were used for Eu and Gd \cite{Nd_choi}. Ferromagnetic ordering was considered following previous studies \cite{jacobs_assessing_2019}. Structural information was obtained from the Inorganic Crystal Structure Database (ICSD) \cite{icsd_2002}, and all crystal structures were fully relaxed until the residual forces on each atom were below 0.01 eV$\cdot$Å$^{-1}$.  A plane-wave energy cutoff of 500 eV and a $k$-point density of more than 7000/atom were used.

To construct an explicit descriptor-activity relationship for CO$_2$ reduction to CH$_4$, we employed a parametric brute-force searching (BFS) algorithm \cite{ghiringhelli_big_2015, you_machine_2020, kim_organized_2016, lee_machine-learning-guided_2021}, which systematically explores candidate equations generated from the given descriptors and their nonlinear transformations to identify the best-performing analytical expression.

In the BFS procedure, each primary descriptor $x$ was transformed using 15 functions:  $1$, $x$, $1/x$, $x^2$, $1/x^2$, $x^3$, $1/x^3$, $\sqrt{x}$, $1/\sqrt{x}$, $e^x$, $1/e^x$, $e^{1/x}$, $1/e^{1/x}$, $\ln(1+x)$, and $1/\ln(1+x)$.
Products of the transformed descriptors were used as compound terms, which were subsequently used in the regression equation
$\hat{A}_{\mathrm{CH_4}}=\theta_0+\sum_{i=1}^{m}\theta_i f_i$, where $\hat{A}_{\mathrm{CH_4}}$ denotes the predicted CH$_4$ activity, $\theta_0$ is the intercept, $\theta_i$ are the regression coefficients, $f_i$ are the compound terms, and $m$ is the number of compound terms included in the equation. In our notation, C denotes the number of compound terms in the regression equation, whereas F denotes the number of primary descriptors used to construct each term. Thus, an $m$C$n$F equation consists of $m$ compound terms, each derived from $n$ primary descriptors through nonlinear transformations.

\begin{figure*}[t]
\centering
\includegraphics[width=0.90\columnwidth]{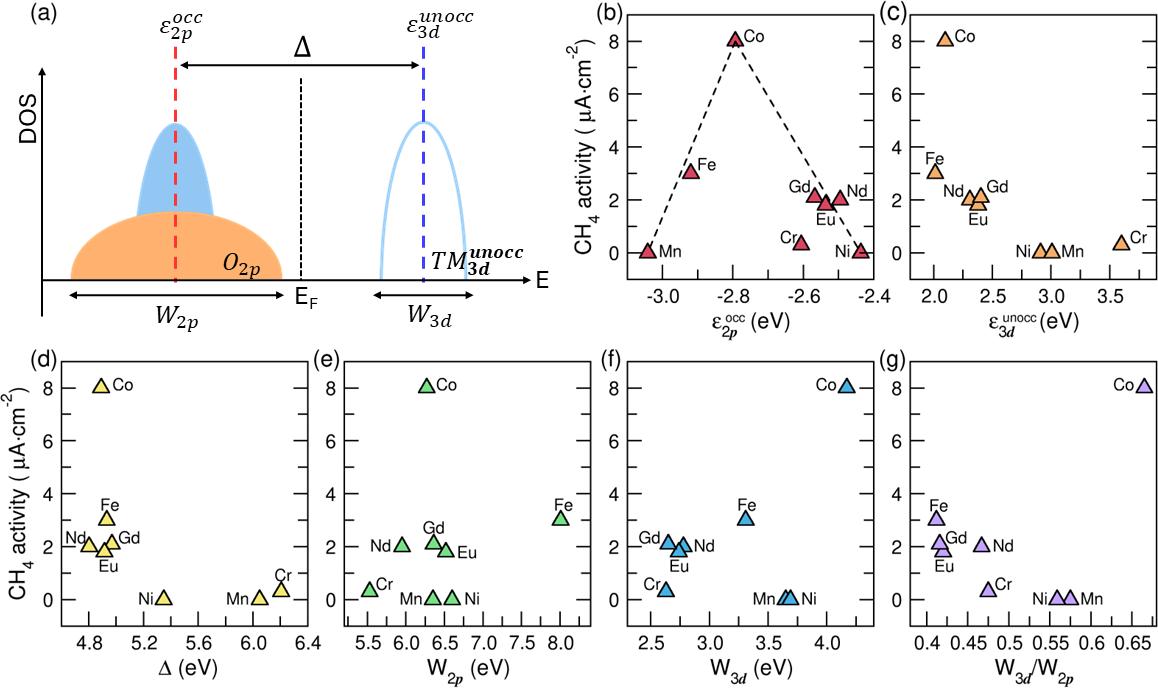}
\caption{\label{fig:Fig2}(a) Schematic illustration of the occupied O $2p$-band center ($\varepsilon_{2p}^{\mathrm{occ}}$), unoccupied transition-metal (TM) $3d$-band center ($\varepsilon_{3d}^{\mathrm{unocc}}$), charge-transfer energy ($\Delta$), occupied O $2p$ bandwidth ($W_{2p}$), and unoccupied TM $3d$ bandwidth ($W_{3d}$). Relationships between the experimentally reported CH$_4$ activity \cite{hwang_2019} and the DFT-calculated descriptors: (b) $\varepsilon_{2p}^{\mathrm{occ}}$, (c) $\varepsilon_{3d}^{\mathrm{unocc}}$, (d) $\Delta$, (e) $W_{2p}$, (f) $W_{3d}$, and (g) $W_{3d}/W_{2p}$. Here, Cr, Mn, Fe, Co, and Ni denote LaCrO$_3$, LaMnO$_3$, LaFeO$_3$, LaCoO$_3$, and LaNiO$_3$, respectively, whereas Nd, Eu, and Gd denote NdCoO$_3$, EuCoO$_3$, and GdCoO$_3$, respectively.}
\end{figure*}

We extracted the key electronic structure descriptors from the calculated density of states (DOS), as schematically illustrated in Fig.~\ref{fig:Fig2}(a). Figure S1 shows the atom-projected DOS for the considered simple perovskites. The O $2p$-band center ($\varepsilon_{2p}^{\mathrm{occ}}$) and the unoccupied TM $3d$-band center ($\varepsilon_{3d}^{\mathrm{unocc}}$) were evaluated from Eqs.~\eqref{eq:O2p_center} and~\eqref{eq:TM3d_center} \cite{pbandcenter_grimaud,yamada_2018}.

\begin{equation}
\varepsilon_{2p}^{\mathrm{occ}} =
\frac{\int_{-\infty}^{E_F} E \,\cdot\, D_{2p}(E)\, dE}
     {\int_{-\infty}^{E_F} D_{2p}(E)\, dE}
\label{eq:O2p_center}
\end{equation}

\begin{equation}
\varepsilon_{3d}^{\mathrm{unocc}} =
\frac{\int_{E_F}^{E_{\max}} E \,\cdot\, D_{3d}(E)\, dE}
     {\int_{E_F}^{E_{\max}} D_{3d}(E)\, dE}
\label{eq:TM3d_center}
\end{equation}

Here, $D_i(E)=D_{i}^{\uparrow}(E) + D_{i}^{\downarrow}(E)$ ($i=2p,\,3d$) denotes the spin-summed projected density of states (PDOS) for the O $2p$ and TM $3d$ orbitals. $E_{\max}$ was chosen as 5 eV, which is sufficient to include the unoccupied TM $3d$ states above the Fermi level.
The charge-transfer energy was calculated as $\Delta=\varepsilon_{3d}^{\mathrm{unocc}}-\varepsilon_{2p}^{\mathrm{occ}}$\cite{hong_CTE, yamada_2018}. The O $2p$ bandwidth ($W_{2p}$) was determined from the occupied states, while the TM $3d$ bandwidth ($W_{3d}$) was evaluated from the unoccupied states relative to the Fermi level. For discontinuously distributed states, the bandwidth was obtained by summing the widths of the separated energy intervals. Table S1 summarizes the calculated descriptor values for the simple perovskites.

Figure~\ref{fig:Fig2}(b)-(g) show the correlations between the experimental CH$_4$ activity reported by Hwang \textit{et al.} \cite{hwang_2019} and the electronic structure descriptors. In Fig.~\ref{fig:Fig2}(b), the O $2p$-band center exhibits a volcano-like relationship with CH$_4$ activity, consistent with the previous study \cite{hwang_2019}. 
In contrast, the unoccupied TM $3d$-band center and the charge-transfer energy ($\Delta$) exhibit more monotonic negative correlations with CH$_4$ activity, as shown in Figs.~\ref{fig:Fig2}(c) and (d).
Because the unoccupied TM $3d$-band center varies over a much wider energy range than the O $2p$-band center, approximately 2.0 eV compared with 0.6 eV, the charge-transfer energy is primarily governed by the variation in the unoccupied TM $3d$-band center.
A smaller charge-transfer energy indicates a reduced separation between the O $2p$ and TM $3d$ manifolds and is generally associated with stronger O $2p$-TM $3d$ hybridization and increased metal-oxygen covalency \cite{hong_CTE}. 
This trend is consistent with previous OER studies on perovskite oxides, where a lower unoccupied TM $3d$-band center and smaller charge-transfer energy were associated with lower OER overpotentials and higher catalytic activity \cite{yamada_2018}.
These results suggest that the relative alignment and hybridization between O $2p$ and TM $3d$ states are also relevant to CO$_2$RR activity.

Beyond the band center-related descriptors, we further examined the bandwidths to account for the energy dispersion of the occupied O $2p$ and unoccupied TM $3d$ states. While the band center characterizes the average energy position of the electronic states, it does not fully describe their distribution. Previous studies have shown that the width and shape of the $d$-state distribution can also influence adsorbate-surface interactions \cite{xin_effects_2014, vojvodic_electronic_2014, ma_machine-learning-augmented_2015}. Accordingly, $W_{2p}$ and $W_{3d}$ were introduced to characterize the bandwidths of the occupied O $2p$ and unoccupied TM $3d$ manifolds, respectively.

As shown in Figs.~\ref{fig:Fig2}(e) and (f), neither $W_{2p}$ nor $W_{3d}$ alone exhibits a clear monotonic relationship with CH$_4$ activity. Nevertheless, except for LaCoO$_3$, higher CH$_4$ activity tends to be associated with a broader occupied O $2p$ manifold and a narrower unoccupied TM $3d$ manifold. This tendency becomes more apparent when the bandwidth ratio $W_{3d}/W_{2p}$ is considered, as shown in Fig.~\ref{fig:Fig2}(g). Lower $W_{3d}/W_{2p}$ values are generally associated with higher CH$_4$ activity. This suggests that $W_{3d}/W_{2p}$, which measures the relative dispersion of the unoccupied TM $3d$ and occupied O $2p$ manifolds, provides a more useful descriptor than either bandwidth considered separately.

\begin{figure}[t]
\centering
\includegraphics[width=0.45\columnwidth]{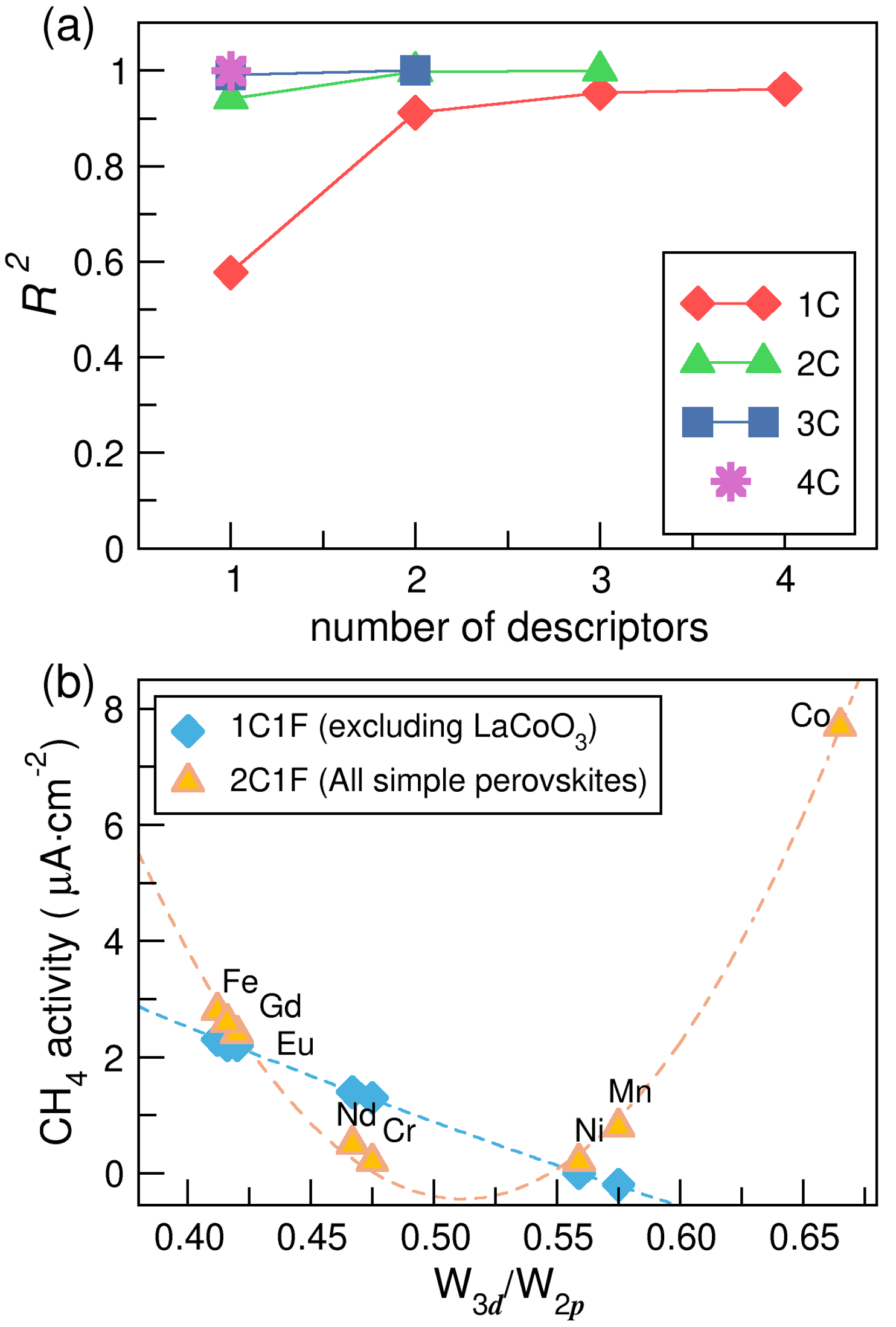}
\caption{\label{fig:Fig3}{Performance of the BFS-derived equations for CH$_4$ activity. (a) $R^2$ of the best-performing BFS equations as a function of the number of descriptors, considering $\varepsilon_{2p}^{\mathrm{occ}}$, $\varepsilon_{3d}^{\mathrm{unocc}}$, $\Delta$, and $W_{3d}/W_{2p}$ for all eight simple perovskites. (b) Comparison of the CH$_4$ activities predicted by the selected 1C1F equation fitted without LaCoO$_3$ and the 2C1F equation fitted using all simple perovskites.}}
\end{figure}

To establish a quantitative relationship between electronic structure and CH$_4$ activity, we applied the parametric BFS approach to four DFT-derived electronic structure descriptors: the occupied O $2p$-band center ($\varepsilon_{2p}^{\mathrm{occ}}$), the unoccupied TM $3d$-band center ($\varepsilon_{3d}^{\mathrm{unocc}}$), the charge-transfer energy ($\Delta$), and the bandwidth ratio ($W_{3d}/W_{2p}$). Figure~\ref{fig:Fig3}(a) shows the performance of the best-performing equations as a function of the number of descriptors, evaluated using the coefficient of determination ($R^2$). In general, the performance improves as the number of descriptors and compound terms increases, with the best 1C2F expression already reaching an $R^2$ of ~0.9. Because the improvement becomes small for more complex equations, we considered up to 4F for 1C, 3F for 2C, 2F for 3C, and 1F for 4C.

Among the equations obtained from the BFS search, we next identified equations that not only showed high fitting performance for the simple-perovskite data but also reproduced the experimentally reported activity trend of La$_2$NiFeO$_6$. Because La$_2$NiFeO$_6$ has been reported to exhibit higher CO$_2$ conversion activity than the related simple perovskites LaNiO$_3$ and LaFeO$_3$ \cite{lim_fundamental_2021}, this trend was used as an additional criterion for equation selection. The simplest equation satisfying this criterion was the 2C1F equation. Notably, despite considering four different electronic structure descriptors in the BFS search, the 10 best-performing 2C1F equations depend only on the bandwidth ratio $W_{3d}/W_{2p}$, rather than combining multiple descriptors. We further examined the behavior of these equations at low $W_{3d}/W_{2p}$ values and excluded equations yielding excessively large or negative activity predictions. The selected equation achieved $R^2=0.9212$ and is given by
\begin{equation}
\hat{A}_{\mathrm{CH_4}}
=314.06\left(\frac{W_{3d}}{W_{2p}}\right)^2 - 601.92\,e^{-\left(W_{2p}/W_{3d}\right)} + 2.68,
\label{eq:2C1F}
\end{equation}
where $\hat{A}_{\mathrm{CH_4}}$ denotes the predicted CH$_4$ activity in $\mu\mathrm{A}\cdot\mathrm{cm}^{-2}$.

To identify a simpler relationship while retaining the main descriptor-activity trend, we further examined 1C1F expressions constructed for the $\varepsilon_{2p}^{\mathrm{occ}}$, $\varepsilon_{3d}^{\mathrm{unocc}}$, $\Delta$, and $W_{3d}/W_{2p}$ electronic structure descriptors after excluding LaCoO$_3$ from the fitting data. Notably, the expressions satisfying the La$_2$NiFeO$_6$ activity criterion again depended solely on $W_{3d}/W_{2p}$. The same descriptor emerged despite the reduced equation complexity and the different fitting conditions. We then examined the behavior of these candidate equations in the low $W_{3d}/W_{2p}$ regime and excluded equations showing divergence in either the predicted activity or its first derivative as $W_{3d}/W_{2p}\rightarrow0$. Among the remaining expressions, the exponential form provided the best fit and was therefore selected. The selected equation achieved $R^2=0.7689$ and is given by
\begin{equation}
\hat{A}_{\mathrm{CH_4}}
=25.45\,e^{-\left(W_{3d}/W_{2p}\right)} - 14.55.
\label{eq:1C1F}
\end{equation}

Figure~\ref{fig:Fig3}(b) compares the CH$_4$ activities predicted by the 2C1F equation fitted to all eight simple perovskites with those predicted by the 1C1F equation obtained after excluding LaCoO$_3$. 
Within the $W_{3d}/W_{2p}$ range spanned by the simple perovskites other than LaCoO$_3$,
the two equations reproduce similar activity trends despite their different complexity. 
This agreement further supports $W_{3d}/W_{2p}$ as a consistent descriptor for capturing the relative CH$_4$ activity trend in the investigated perovskites.
Accordingly, the simpler 1C1F expression was adopted for the subsequent prediction of CH$_4$ activities in double perovskites.

\begin{figure}[t]
\centering
\includegraphics[width=0.6\columnwidth]{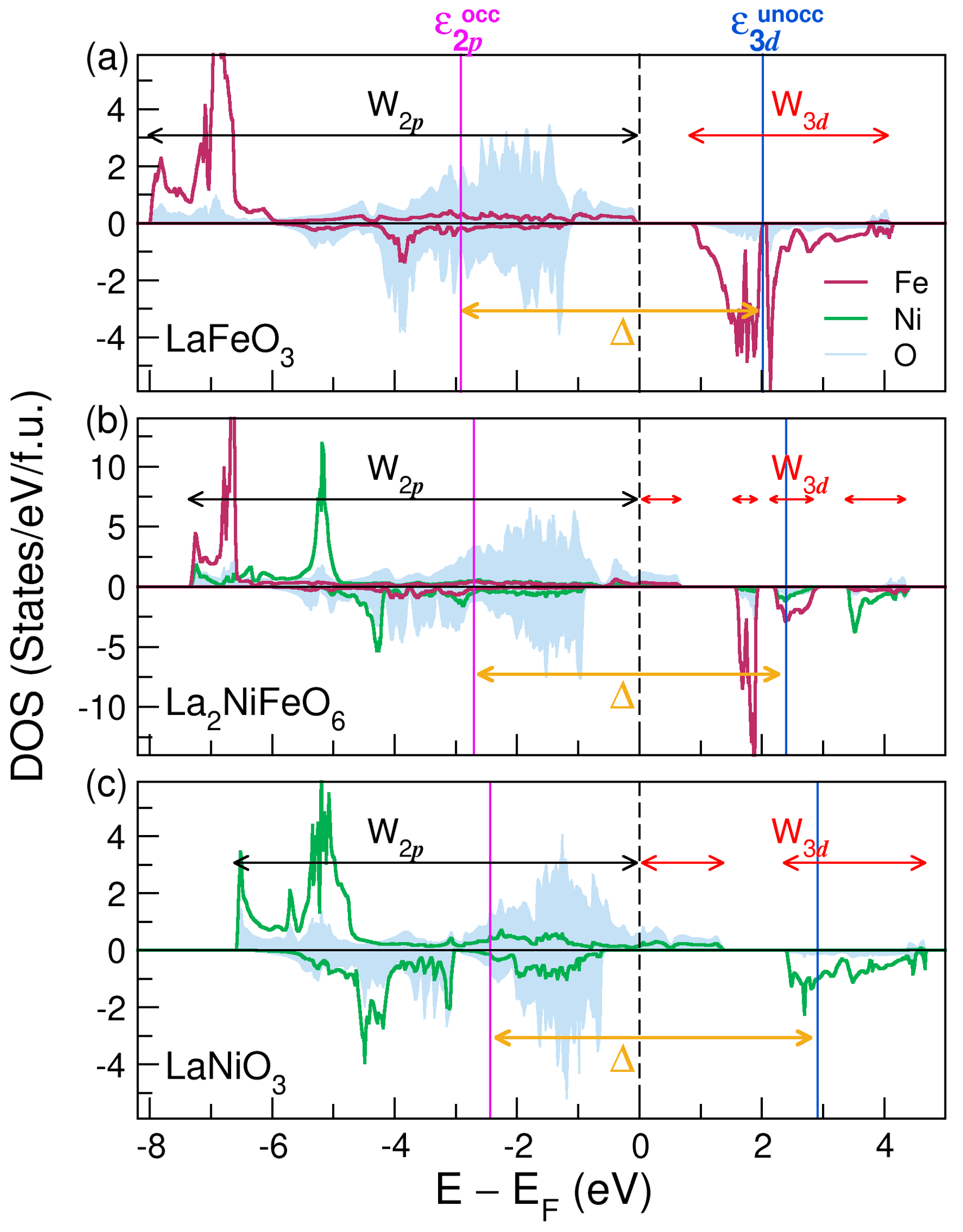}
\caption{\label{fig:Fig4}
Atom-projected densities of states (DOS) of (a) LaFeO$_3$, (b) La$_2$NiFeO$_6$, and (c) LaNiO$_3$. The vertical magenta and blue lines indicate the occupied O $2p$-band center ($\varepsilon_{2p}^{\mathrm{occ}}$) and the unoccupied TM $3d$-band center ($\varepsilon_{3d}^{\mathrm{unocc}}$), respectively. The charge-transfer energy ($\Delta$) and the bandwidths, $W_{2p}$ and $W_{3d}$, are denoted by horizontal arrows.
}
\end{figure}

To understand why $W_{3d}/W_{2p}$ was repeatedly selected in the BFS expressions, we compared the electronic structures of LaFeO$_3$, La$_2$NiFeO$_6$, and LaNiO$_3$. Figures~\ref{fig:Fig4}(a--c) show the spin-resolved atom-projected DOS of LaFeO$_3$, La$_2$NiFeO$_6$, and LaNiO$_3$, respectively. Most of the electronic structure descriptors vary monotonically across these three compounds. Specifically, the occupied O $2p$-band center shifts toward the Fermi level while the unoccupied TM $3d$-band center shifts to higher energy. The charge-transfer energy $\Delta$ increases and $W_{2p}$ decreases from LaFeO$_3$ through La$_2$NiFeO$_6$ to LaNiO$_3$. Accordingly, the corresponding descriptor values of La$_2$NiFeO$_6$ generally lie between those of LaFeO$_3$ and LaNiO$_3$. Such intermediate changes alone do not account for the experimentally reported higher activity of La$_2$NiFeO$_6$ relative to both simple perovskites.

In contrast, the unoccupied TM $3d$ states of La$_2$NiFeO$_6$ are distributed over a narrower energy range than those of LaFeO$_3$ and LaNiO$_3$, resulting in a smaller $W_{3d}$. Consequently, the reduced $W_{3d}/W_{2p}$ ratio of La$_2$NiFeO$_6$ deviates from the monotonic variation between LaFeO$_3$ and LaNiO$_3$, allowing the equation to reproduce its enhanced CO$_2$ conversion activity. The reduction in $W_{3d}$ may be associated with the different energy levels of the Ni and Fe $3d$ states, which can weaken the effective intersite coupling and reduce the dispersion of the unoccupied TM $3d$ bands. Thus, $W_{3d}/W_{2p}$ provides an electronic structure basis for describing the enhanced activity of La$_2$NiFeO$_6$, while also capturing the activity trend among the simple perovskites.

Finally, we applied the selected 1C1F equation to screen 24 La-based double perovskites for CH$_4$ activity. Among them, 21 compositions were constructed from all pairwise combinations of the $3d$ transition metals from Ti to Ni. La$_2$CoAlO$_6$, La$_2$CoGaO$_6$, and La$_2$CoCuO$_6$ were additionally included because B-site substitution of LaCoO$_3$ with Al, Ga, and Cu has previously been reported to enhance CO$_2$ conversion performance \cite{goto_enhanced_2024,dong_transition_2022}. The calculated electronic structure descriptors of the investigated double perovskites are summarized in Table S2.
\begin{figure}[t]
\centering
\includegraphics[width=0.55\columnwidth]{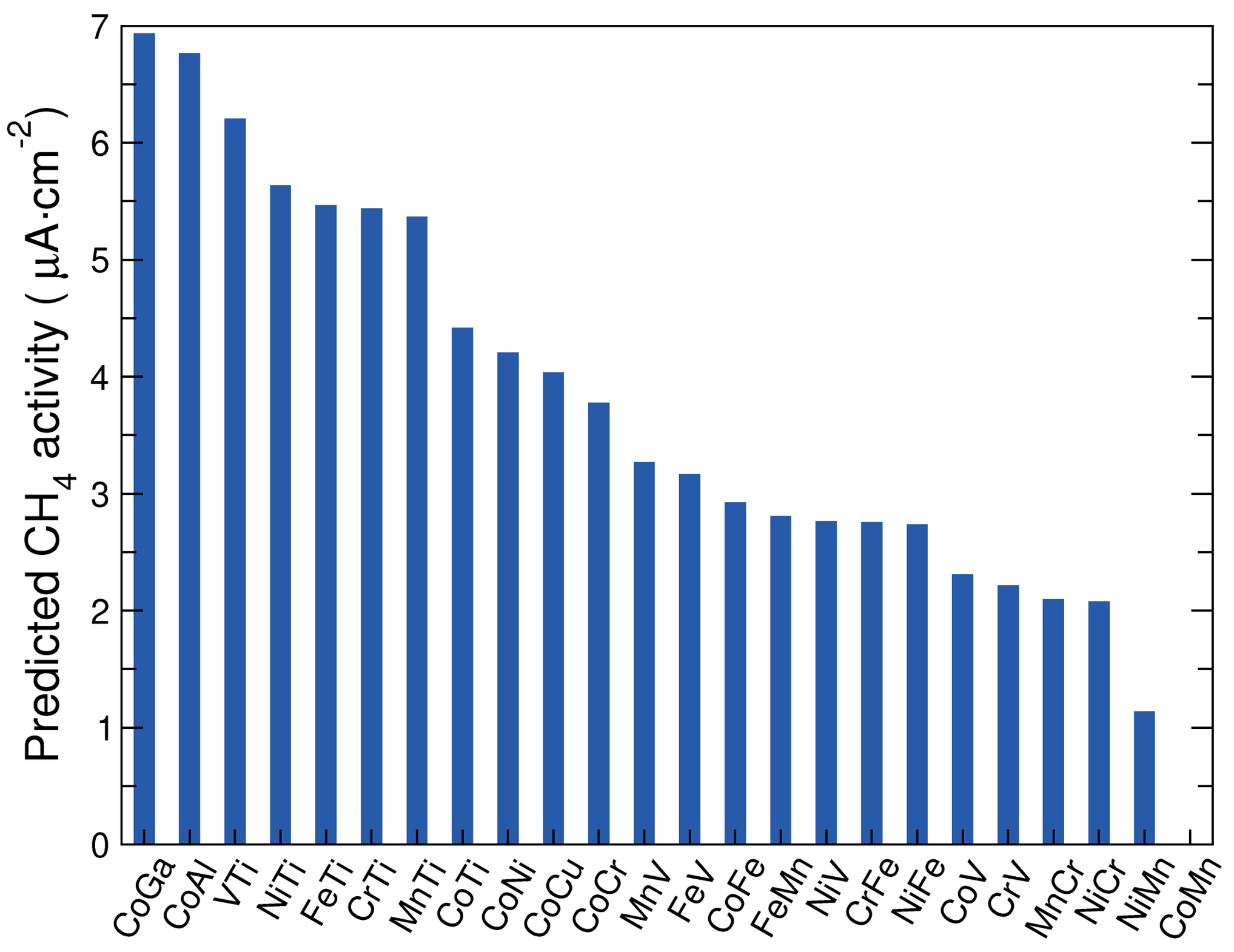}
\caption{\label{fig:Fig5}Predicted CH$_4$ activities of double perovskites, arranged in descending order. The abbreviated labels indicate B-site combinations; for example, CoGa represents La$_2$CoGaO$_6$.}
\end{figure}

Figure~\ref{fig:Fig5} shows the predicted CH$_4$ activities of the 24 double perovskites in descending order. La$_2$CoGaO$_6$ and La$_2$CoAlO$_6$ exhibit the highest predicted activities among the examined compositions, while La$_2$CoCuO$_6$ and La$_2$CoNiO$_6$ also show moderately high predicted activities. The relatively favorable ranking of these Co-based compositions is qualitatively consistent with previous reports showing enhanced CO$_2$ conversion upon B-site substitution of LaCoO$_3$ \cite{goto_enhanced_2024,dong_transition_2022}. Several Ti-containing double perovskites also appear in the higher-activity region, whereas Mn-containing compositions tend to show relatively low predicted activities. The favorable trend predicted for the Ti-containing compositions is also qualitatively consistent with a recent experimental report showing enhanced CO$_2$ electrolysis performance upon Ti incorporation in a related perovskite oxide\cite{zhang_synergistic_2026}.

The DOS analysis provides a consistent electronic structure picture for these trends, as shown in Figure S2. La$_2$CoGaO$_6$ and La$_2$CoAlO$_6$ exhibit broad occupied O $2p$ states together with narrow unoccupied TM $3d$ states, resulting in small $W_{3d}/W_{2p}$ ratios. Pronounced narrowing of the unoccupied TM $3d$ states is also observed for Ti-containing double perovskites, despite their comparatively smaller $W_{2p}$. In contrast, Mn-containing compositions near the low-activity end, such as La$_2$NiMnO$_6$ and La$_2$CoMnO$_6$, exhibit broader unoccupied TM $3d$ distributions and correspondingly larger $W_{3d}/W_{2p}$ ratios.

In conclusion, we investigated electronic structure descriptors associated with CO$_2$ reduction to CH$_4$ in perovskite oxides. Among the descriptors considered, the unoccupied TM $3d$-band center, charge-transfer energy, and $W_{3d}/W_{2p}$ showed negative correlations with CH$_4$ activity. Using the BFS approach, we searched for explicit fitting expressions and further used the experimentally reported enhanced activity of La$_2$NiFeO$_6$ relative to LaNiO$_3$ and LaFeO$_3$ as an additional criterion for equation selection. Notably, the selected equations repeatedly depended on $W_{3d}/W_{2p}$.
The DOS comparison further showed that the $W_{3d}/W_{2p}$ ratio of La$_2$NiFeO$_6$ does not simply take an intermediate value between those of LaFeO$_3$ and LaNiO$_3$. Instead, the narrower distribution of its unoccupied TM $3d$ states leads to a smaller ratio, providing an electronic structure basis for its enhanced activity. The selected 1C1F equation was then applied to screen 24 La-based double perovskites. La$_2$CoGaO$_6$ and La$_2$CoAlO$_6$ exhibited the highest predicted CH$_4$ activities. Ti-containing compositions also showed relatively high predicted activities, while Mn-containing compositions tended to show lower values. These trends were consistent with the corresponding DOS characteristics.
Overall, our results suggest that $W_{3d}/W_{2p}$ is a useful descriptor for relating electronic structure variations to CH$_4$ activity in simple and double perovskites and for identifying candidate compositions for further investigation.

\section*{ASSOCIATED CONTENT}
\section*{Supporting Information}
The Supporting Information is available free of charge at [DOI].

\begin{itemize}
  \item Spin-resolved atom-projected densities of states for the simple and selected double perovskites, and calculated electronic structure descriptors for the simple and double perovskites (PDF).
\end{itemize}

\section*{AUTHOR INFORMATION}
\subsection*{Corresponding Author}
\hspace*{2em} \textbf{Sooran Kim} -- Department of Physics Education, Kyungpook National University, Daegu 41566, South Korea; KNU G-LAMP Project Group, KNU Institute of Basic Sciences, Kyungpook National University, Daegu 41566, South Korea
 \date{Email: sooran@knu.ac.kr}
\subsection*{Authors}
\hspace*{2em} \textbf{Hyo-sun Jin} -- Department of Physics Education, Kyungpook National University, Daegu 41566, South Korea; KNU G-LAMP Project Group, KNU Institute of Basic Sciences, Kyungpook National University, Daegu 41566, South Korea\\
\hspace*{2em} \textbf{Jiyeon Kim} -- Department of Physics Education, Kyungpook National University, Daegu 41566, South Korea
\subsection*{Notes}
The authors declare no competing financial interest. 

\section*{ACKNOWLEDGMENTS}
We thank Bongjae Kim for helpful discussions.
This work was supported by the National Research Foundation of Korea (NRF) (No. RS-2025-00557388, RS-2026-25542082) and the KISTI supercomputing center (No. KSC-2025-CRE-0461). This work was also supported by Global-Learning \& Academic research institution for Master’s· PhD students and Postdocs (LAMP) Program of the NRF grant funded by the Ministry of Education (No. RS-2023-00301914).

%%%%%%%%%%%%%%%%%%%%%%%%%%%%%%%%%%%%%%%%%%%%%%%%%%%%%%%%%%%%%%%%%%%%%
%% If you are using classical BibTeX rather than biblatex,
%% remove the \printbibliography and uncomment the \bibliograpy one
%%%%%%%%%%%%%%%%%%%%%%%%%%%%%%%%%%%%%%%%%%%%%%%%%%%%%%%%%%%%%%%%%%%%%
%\printbibliography
\bibliographystyle{chem-acs}
\renewcommand{\refname}{REFERENCES}
\bibliography{ref}
\end{document}